\documentstyle[epsfig,12pt]{article}
\newcommand\slv{v\kern-5pt\raise1pt\hbox{$\scriptstyle/$}\kern1pt}

\begin{document}
\begin{flushright}
WUE-ITP-97-050\\
SPhT-T97/139
\end{flushright}
\vspace{0.5cm}

\begin{center}
{\Large \bf $|V_{ub}|$ and Perturbative QCD Effects }\\[1mm]
{\Large \bf in the $B\to\pi$ Transition Form Factor}
\footnote{This work is supported by
BMBF under contract number 05 7WZ91P(0);\\ 
Talk given by O. Yakovlev at IV International Workshop on Progress in
Heavy
          Quark Physics, Rostock, September 20-22.}\\
\vspace{2mm}
{\sc \bf 
S.~Weinzierl$^a$ and O.~Yakovlev$^{b,c}$}\\[3mm]
\end{center}
{\small 
$^a$ Centre d'Etudes de Saclay,
$^b$ Universit\"at W\"urzburg,
$^c$ Budker INP Novosibirsk.
}\\[2mm]
{\small {\bf Abstract:} We report on recent improvements for the
$B\to\pi$ form factor. 
The updated value of $|V_{ub}|$ is presented.}\\[2mm]
{\bf 1. Motivation.}  The semileptonic decay $B\to\pi l\nu_l $ 
is one of the most important reactions for the 
determination of the CKM parameter 
$V_{ub}$. However, in order to extract $V_{ub}$  
from data one needs an accurate 
theoretical calculation of    
 the hadronic matrix element
\begin{eqnarray}\label{formdef}
\langle\pi (q)|\bar u\gamma_{\mu}b |B(p+q)\rangle
=2f^+(p^2)q_{\mu}+(f^+(p^2)+
f^-(p^2))p_{\mu},
\end {eqnarray}
where $p+q$, $q$ and $p$ denote the $B$ and $\pi$ four-momenta 
and the momentum transfer, respectively, and $f^{\pm}$ are two 
independent form factors. 
A very reliable approach 
to calculate $f^{\pm}$ in the framework of QCD  
is provided by the operator
product expansion (OPE) on the light-cone [1,2,3] in 
combination with QCD sum rule techniques.
The sum rule for the form factor $f^{+}(p^2)$ has been 
obtained in the leading order (LO) in $\alpha_s$ 
in [4,5] taking into account twist 2, 3 and 4 operators.
The most important missing elements of these calculations 
are the perturbative QCD corrections to the correlation function. 
Here we report on a results of the calculation
of the $O(\alpha_s)$ correction to $f^+$ [13,14] 
which eliminates one of the main 
uncertainty in the sum rule results. 

{\bf 2. Sum rule.} The general idea of the sum rule 
method is to consider the correlation
function of two heavy-light currents,
 \begin{eqnarray}
 F_{\mu}(p,q)&=&i\int dxe^{ip\cdot x}
 \langle\pi (q)|T\{\bar u (x)\gamma_\mu b(x) , m_b\bar b(0)i\gamma_5 
 d(0)\}|0\rangle
\label{corr} 
\end{eqnarray}
which can be calculated in the region $(p+q)^2<0 $ and  
$p^2 < m_b^2 - O($1GeV$^2)$ using OPE near the light-cone, 
i.e. at $x^2 \simeq 0$. 
In this note we focus on 
the leading twist 2 contribution. The sum rule for $f^+$
in LO in $\alpha_s$ is given by
\begin{eqnarray}
f_Bf^+(p^2)=\frac{f_\pi m_b^2}{2m^2_B}\int^{s_0}_{m_b^2} \varphi_\pi  (u_0)
e^{\frac{m^2_B-s}{M^2}}ds +\mbox{higher twists}~,
\label{sr}
\end{eqnarray} 
Here $\varphi_\pi (u)$ is the 
pion wave function, $m_b,M_B$ are  the masses of the heavy 
quark and meson, $M^2$ is the Borel parameter, $s_0$ is  the threshold of the 
continuum, $f_\pi=132$MeV, $u_0=\frac{m_b^2-p^2}{s-p^2}$.
The calculation  has several aspects which are worth pointing out.
Firstly, the sum rule is actually derived for the product 
$f_Bf^+$, $f_B$ being  
the $B$ meson decay constant defined by
$ 
\langle B|\bar bi\gamma_5 d|0 \rangle =m^2_Bf_B/m_b~.
$
The form factor $f^+$ itself is then obtained by dividing out 
$f_B$ taking  the value determined from  
the corresponding two-point QCD sum rule. 
In previous estimates, 
the $O(\alpha_s)$ correction to $f_B$ 
was thereby ignored for consistency because of the lack of the 
$O(\alpha_s)$ correction to $f_Bf^+$. Our calculation 
now allows to take 
into account the correction to $f_B$ which
is known to be sizeable.  
Secondly, knowing the $O(\alpha_s)$ corrections, also 
the heavy quark mass entering the sum rule can be 
properly defined.
The calculation for a finite quark 
mass is new and will have numerous applications.
 
{\bf 3. QCD correction.} The correlator can be written as a convolution 
of a hard amplitude $T(p^2,(p+q)^2,u)$ calculable within 
perturbation theory, with the pion wave function  
$\varphi_{\pi}(u)$ containing the long-distance effects:
\begin{eqnarray}\label{represent}
F(p^2,(p+q)^2)=-f_\pi\int^1_0 du \varphi_\pi (u) T(p^2,(p+q)^2,u).
\end{eqnarray}
The evolution of the light-cone wave function $\varphi_\pi(u)$ 
is controlled by the Brodsky-Lepage equation [2]
\begin{equation}
d\varphi_\pi (u,\mu)/d\ln \mu = \int^1_0 d\omega V(u,\omega)
\varphi_\pi(\omega,\mu) 
\label{BLL}
\end{equation}
The first step is to calculate the 
$O(\alpha_s)$ correction to the hard amplitude $T$. 
The calculation is performed 
in general covariant gauge in order 
to have a possibility to check the gauge invariance of the result. 
Both the ultraviolet (UV)  and infrared
divergences are regularized by 
dimensional regularization and renormalized 
in the $\overline{MS}$ scheme 
with  totally anticommuting $\gamma_5$.  
This choice is motivated by the fact that the same 
scheme is used in the 
calculation of the NLO evolution kernel of the wave function 
$\varphi_\pi(u)$ [6]. 
After UV renormalization, IR factorization and reexpressing of the $\overline{MS}$
mass by the pole mass, we have obtained
$$
T(r_1,r_2,u,\mu) =  
\frac1{\rho-1}
+\frac{\alpha_s(\mu)C_F}{4\pi}
 \Bigg\{\frac1{\rho-1}( -4+3
\ln \frac{m_b^{*2}}{\mu^2}) 
+\frac{2}{\rho-1} \left[ 
     2 G\left(\rho\right) - G\left(r_1\right) - G\left(r_2\right)
 \right] 
$$
$$       +\frac{2}{(r_1-r_2)^2} \left(
             \frac{1-r_2}{u} \left[ G\left(\rho\right) -
 G\left(r_1\right)\right]
           + \frac{1-r_1}{1-u} \left[ G\left(\rho\right) - 
G\left(r_2\right)\right] \right) 
$$
$$
       +\frac{\rho+(1-\rho)\ln\left(1-\rho\right)}{\rho^2}
+\frac{2}{\rho-1} \frac{(1-r_2)\ln\left(1-r_2\right)}{r_2} 
           -\frac{2}{\rho-1}
$$
\begin{equation}
       -\frac{2}{(1-u)(r_1-r_2)}
             \left( \frac{ (1-\rho) \ln\left(1-\rho\right)}{\rho}
                  - \frac{ (1-r_2) \ln\left(1-r_2\right)}{r_2}  \right)
    \Bigg\} ~.
\label{result}
\end{equation}
We used convenient dimensionless 
variables $r_1 = p^2/m_b^2$ and $r_2=(p+q)^2/m_b^2$ and 
\begin{eqnarray}
&&\Delta = \frac{2}{4-d}-\gamma_E+\ln (4\pi ),\quad \quad 
\rho = r_1 + u (r_2-r_1), \\
G \left(\rho \right) 
& = & \mbox{Li}_2(\rho) + \ln^2(1-\rho) +\ln(1-\rho)
\left(\ln\frac{m_b^2}{\mu^2}
 +1 \right),\nonumber 
\end{eqnarray}
$\mbox{Li}_2(x)=-\int\limits^x_0\frac{dt}t \ln(1-t)$ being the Spence function.
The UV renormalization scale 
and the factorization
scale of the collinear (COL) divergences  are taken  
to be equal and denoted by $\mu$. 
 As an additional check on 
the origin of the various divergent
terms we have performed additional explicit calculations.
In particular, we have used mass regularization 
by giving the light quarks a small but finite mass, 
and momentum regularization keeping the light quarks off mass shell.



{\bf 4. Numerical results.} 
The next step is to determine the decay constant $f_B$ and 
the pion wave function  $\varphi_\pi(u,\mu)$ in NLO. For that purpose  
we have analyzed the two-point sum rule
for $f_B$ obtained from    
the renormalization-group-invariant 
correlation function\\ 
$m_b^2\langle 0\mid T\{J_5^+(x)J_5(0)\} \mid 0 \rangle$   
in $O(\alpha_s)$ [7]. For the  running coupling constant 
we use the two-loop expression 
with $N_f=4$ and  
$\Lambda^{(4)}=234$ MeV [8]
corresponding  to $\alpha_s(M_Z)= 0.112$. 
For $\mu^2$ we take the value   $\mu^2_B= m_B^2-m_b^{*2}$.  
corresponding to the average virtuality 
of the correlation function.
With this choice the following 
correlated results are extracted from the two-point
sum rule: 
\begin{eqnarray}
\label{fbb}
 f_{B}=180\pm 30\quad\mbox{MeV} \qquad 
m_b^*=4.7\mp0.1\quad\mbox{GeV},\qquad 
s_0=35\pm 2\quad\mbox{GeV}^2.
\end{eqnarray} 
In the following, we adopt the central values in the above intervals.
Note that without $O(\alpha_s)$ correction 
one obtains $f_B = 140 \pm 30 $ MeV.   
The remaining parameters entering the sum rules are 
directly measured: $m_B=5.279 $ GeV and $f_{\pi}=132$ MeV.

For the wave function $\varphi_\pi$ 
we adopt the ansatz suggested in [9]:
\begin{eqnarray}  
\varphi_\pi(u,\mu_0)=\Psi_0(u)+a_2(\mu_0) \Psi_2(u) 
+a_4(\mu_0) \Psi_4(u),
\end{eqnarray}
where 
$ \quad \Psi_{n}(u) =  6u (1-u) C_{n}^{3/2}(2 u -1)$. 
The coefficients $a_2(\mu_0)=2/3$ and $a_4(\mu_0)=0.43$
at the scale $\mu_0=500$ MeV 
have been extracted [9] from a two-point QCD sum rule  
for the moments of $\varphi_\pi(u)$ [1]. 

 Now we are ready to perform a numerical analysis of the   
sum rule. In Fig. 1, the product 
$f_Bf^+(0)$ is plotted as a function of the Borel parameter $M^2$.    
The $O(\alpha_s)$ correction turns out to be large, 
between 30\%  and  35\% , and stable 
under variation of $M^2$. 
 Fig. 2 shows the momentum 
dependence of the form factor $f^+(p^2)$  
in the region $0<p^2<15\div17$ GeV$^2$
 for $M^2=10$ GeV$^2$, where the sum rule
 is expected to be valid.
Note the almost complete cancellation of the  
NLO correction in $f^+$.  
Finally, it is interesting to compare the $\mu$ dependence
in LO and NLO (Fig.3). The very
mild $\mu$ -dependence in LO results from the evolution
of the wave function. In NLO, the $\mu$-dependence is 
stronger than in LO but similar to the $\mu$-dependence 
of $f_B$. As a result, the residual scale dependence 
of $f^+$ is again mild.

{\bf 5. Application.} 
The above results refer to the leading twist 2  
approximation. The thorough numerical analysis of the NLO sum rules
have been performed in [10] taking into account LO twist 3 and 
4 contributions. Here we give preliminary numbers. 
The final result for the form factor 
$f^+(r), r=\frac{p^2}{m_B^2}$ can be approximate by the function [10] 
(see also [14,15])
$$
f^+(r)=\frac{f^+(0)}{1-ar+br^2}.
$$
with $a=1.5, b=0.52$ and  
$$
f^+(0) = 0.27\pm 0.02 \pm 0.02 ~.
$$
The first uncertainty is connected with  the unknown perturbative 
corrections  to the twist-2 ($O(\alpha_s^2)$)and twist-3 ($O(\alpha_s)$) 
contributions and  the second one is connected 
with  the wave functions. 
This value is $10\% $ lower than the $LO$ estimate 
$f^+(0) = 0.30 $ obtained in [4,11].

Integrating over momentum one obtains the decay width [10]
$$
\Gamma (B^0\to\pi^-e^+\nu_e) =(7.5 \pm 2)|V_{ub}|^2 ps^{-1}.
$$
And finally, using the current CLEO number for the 
$Br(B^0\to\pi l\nu)=(1.8\pm 0.4)
\cdot10^{-4}$ [12] and the world average of the $B^0$ lifetime 
$\tau_{B^0}=(1.56\pm 0.06)$ ps [8] one obtains
$$
|V_{ub}|^{B\to\pi}=0.0039\pm 0.0005_{exp}\pm 0.0005_{th}.
$$
where we indicate the theoretical and experimental uncertainty.

{\bf 6. Outlook.} Here we stress that the light cone sum rule 
gives a reliable estimation for the $f^+(p^2)$ form factor. The present 
accuracy of the result is estimated to be $15-20\%$ and 
can be improved up to $10\% $ by including  
the unknown perturbative $O(\alpha_s^2)$ correction 
to twist-2 and the $O(\alpha_s)$ correction to 
the twist-3 contributions and by more accurate extraction of the 
pion wave functions from the data. \\
{\bf 7. Acknowledgements.} 
We are grateful to R. R\"uckl, A. Khodjamirian, Ch. Winhart for the 
collaboration. We would like to thank P. Ball, V. Braun, 
A. Grozin, M. Neubert, K. Melnikov and A. Vainshtein for useful discussions.
\newpage 
{ \small
{\bf Refernces}\\
1. V.L. Chernyak, A.R. Zhitnitsky,
JETP Lett.  {\bf {25}} (1977) 510;
  Sov. J. Nucl. Phys. {\bf {31}} (1980) 544; 
  Phys. Rep.  {\bf {112}} (1984) 173. \\
2. A.V. Efremov, A.V. Radyushkin,
  Phys. Lett.  {\bf {B94}} (1980) 245;
T.M.F.{\bf {42}} (1980) 147.\\
3. G.P. Lepage,S.J. Brodsky,
  Phys. Lett.  {\bf {B87}} (1979) 359;
  Phys. Rev.  {\bf {D22}} (1980) 2157.\\
4. V.M. Belyaev, A. Khodjamirian and  R. R\"uckl, 
Z. Phys. {\bf C 60} (1993) 349.\\  
5. V.M. Belyaev, V.M.~Braun, 
A. Khodjamirian and  R. R\"uckl, 
Phys. Rev {\bf D 51} (1995) 6177.\\ 
6. F.M. Dittes and A.V. Radyushkin, Phys. Lett. {\bf B134} (1984) 359; 
M.H. Sarmadi, Phys. Lett. {\bf B143} (1984) 471; 
S.V. Mikhailov and A.V. Radyushkin, Nucl. Phys. {\bf B254}  (1985) 89.\\
7. D.J. Broadhurst and S.C. Generalis,
OUT-4102-8/R (1981); 
D.J. Broadhurst, Phys. Lett. {\bf B101} (1981) 423; 
T.M. Aliev and V.I. Eletsky, Sov. J. Nucl. Phys. {\bf 38} (1983) 936.\\
8. Particle Data Group, Phys. Rev. {\bf D54} (1996) 1.\\ 
9. V.M. Braun and I.E. Filyanov, Z. Phys. {\bf C44} (1989) 157.\\
10. A. Khodjamirian, R. R\"uckl, S. Weinzierl, Ch. Winhart and O. Yakovlev,
to be published.\\
11. A. Khodjamirian and R. R\"uckl, in {\it 
``Continuous Advances in QCD 1996''},\\ edited 
by M.I. Polikarpov ( World Scientific, Singapore, 1996), pp. 75-83.\\ 
12. L. Gibbons, {\it 7-th Int. Symp. on Heavy 
Quark Physics, Santa Barbara, USA, July 1997.}\\
13. A. Khodjamirian, R. R\"uckl, S. Weinzierl and O. Yakovlev,
Phys. Lett. B410 (1997) 275.\\
14. E. Bagan, P. Ball, V. Braun, hep-ph/9709243.\\
15.  A. Khodjamirian and R. R\"uckl, to appear in Heavy Flavours, 
2nd edition, eds. A.J. Buras and M. Lindner (World
Scientific, Singapore).}
\newpage


\begin{figure}
\begin{center}
\epsfig{figure=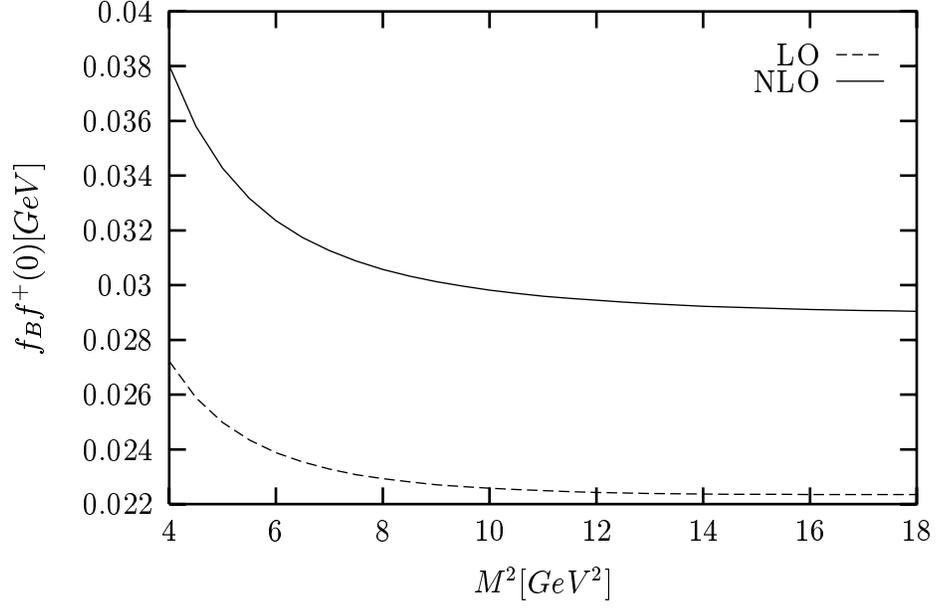}
\end{center}
\caption{\small Light-cone sum rule estimate for $f_Bf^+(0)$ in leading twist 2
approximation as a function of the Borel parameter $M^2$: NLO (solid ) in
comparison to LO (dashed).}
\end{figure}
\begin{figure}
\begin{center}
\epsfig{figure=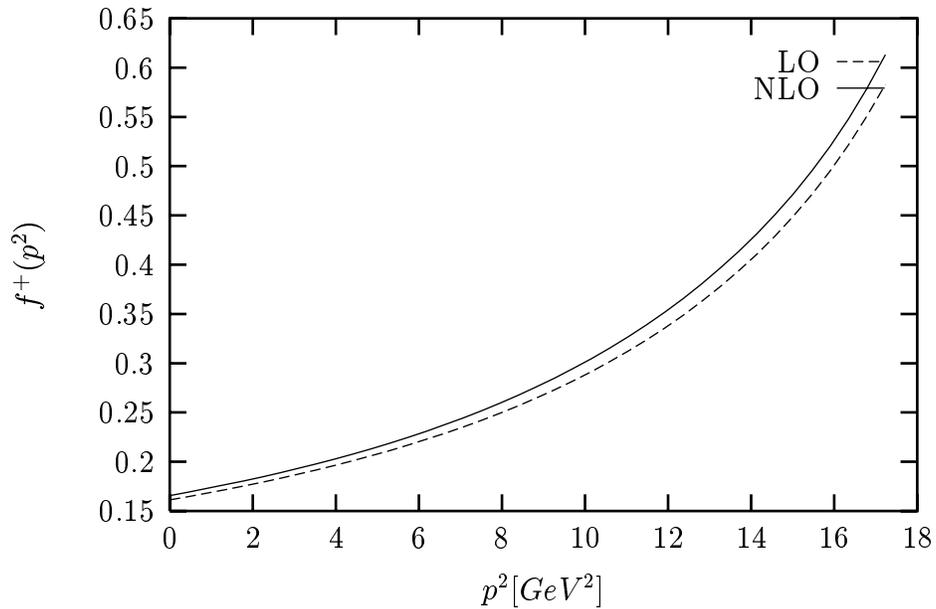}
\end{center}
\caption{\small Momentum dependence of the form factor $f^+(p^2)$ in leading twist 2
approximation: LO (dashed) in comparison to NLO (solid).}
\end{figure}

\begin{figure}
\begin{center}
\epsfig{figure=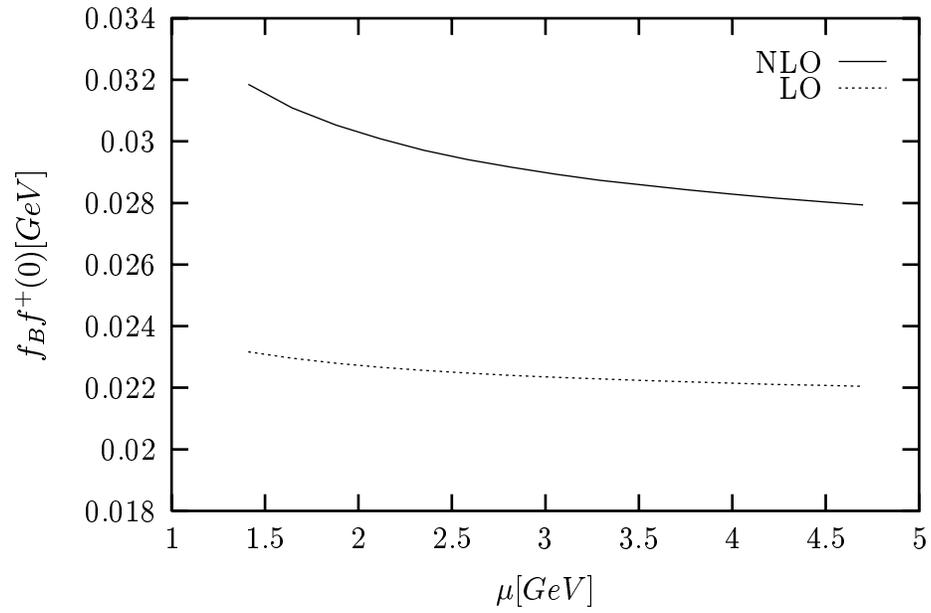}
\end{center}
\caption{\small Scale dependence of the light-cone sum rule estimate of 
$f_Bf^+(0)$ in leading twist 2 approximation: NLO (solid) in comparison to LO
(dotted).}
\end{figure}

\end{document}